\documentclass[lettersize,journal]{IEEEtran}
\usepackage{amsmath,amsfonts}
\usepackage{array}
\usepackage[caption=false,font=normalsize,labelfont=sf,textfont=sf]{subfig}
\usepackage{textcomp}
\usepackage{stfloats}
\usepackage{url}
\usepackage{verbatim}
\usepackage{graphicx}
\def\BibTeX{{\rm B\kern-.05em{\sc i\kern-.025em b}\kern-.08em
    T\kern-.1667em\lower.7ex\hbox{E}\kern-.125emX}}
\usepackage{balance}

\usepackage{cite}
\usepackage{url}
\usepackage[backref]{hyperref}
\usepackage{pifont}

\usepackage[ruled,vlined]{algorithm2e}

\usepackage{tikz}
\usepackage{etoolbox}
\usepackage{enumitem}
\newcommand*{\circled}[1]{\lower.7ex\hbox{\tikz\draw (0pt, 0pt)%
    circle (.5em) node {\makebox[1em][c]{\small #1}};}}
\robustify{\circled}

\begin{document}

\title{A Trustworthy and Consistent Blockchain Oracle Scheme for Industrial Internet of Things}

\author{Peng Liu, Youquan Xian, Chuanjian Yao, Peng Wang, Li-e Wang, Xianxian Li

\thanks{The research was supported in part by  the Guangxi Science and Technology Major Project (No.AA22068070), the National Natural Science Foundation of China (Nos.62166004, U21A20474,62262003), the Key Lab of Education Blockchain and Intelligent Technology, the Center for Applied Mathematics of Guangxi, the Guangxi "Bagui Scholar" Teams for Innovation and Research Project, the Guangxi Talent Highland Project of Big Data Intelligence and Application, the Guangxi Collaborative Center of Multisource Information Integration and Intelligent Processing.}

\thanks{Peng Liu, Youquan Xian, Chuanjian Yao, Peng Wang, Li-e Wang and Xianxian Li are affiliated with both the Key Lab of Education Blockchain and Intelligent Technology, Ministry of Education, and the Guangxi Key Lab of Multi-Source Information Mining and Security at Guangxi Normal University, Guilin, 541004, China. (e-mail: liupeng@gxnu.edu.cn; xianyouquan@stu.gxnu.edu.cn; yaochuanjian@stu.gxnu.edu.cn; wangp@gxnu.edu.cn; wanglie@gxnu.edu.cn; lixx@gxnu.edu.cn), Xianxian Li is the corresponding author.}

\thanks{Manuscript created October, 2022.}}

\markboth{Journal of \LaTeX\ Class Files,~Vol.~18, No.~9, September~2020}%
{How to Use the IEEEtran \LaTeX \ Templates}

\maketitle

\begin{abstract}
Blockchain provides decentralization and trustlessness features for the Industrial Internet of Things (IIoT), which expands the application scenarios of IIoT. To address the problem that the blockchain cannot actively obtain off-chain data, the blockchain oracle is proposed as a bridge between the blockchain and external data. However, the existing oracle schemes are difficult to solve the problem of low quality of service caused by frequent data changes and heterogeneous devices in IIoT, and the current oracle node selection schemes are difficult to balance security and quality of service. To tackle these problems, this paper proposes a secure and reliable oracle scheme that can obtain high-quality off-chain data. Specifically, we first design an oracle node selection algorithm based on Verifiable Random Function (VRF) and reputation mechanism to securely select high-quality nodes. Second, we propose a data filtering algorithm based on a sliding window to further improve the consistency of the collected data. We verify the security of the proposed scheme through security analysis. The experimental results show that the proposed scheme can effectively improve the service quality of the oracle.
\end{abstract}


\section{Introduction}

\IEEEPARstart{B}{lockchain} is a distributed technology for recording and storing data. It has gained significant attention in recent years due to its technical features, such as decentralization, trustlessness, and traceability. Based on these features, blockchain can act as a trusted intermediary between enterprises in IIoT, creating a trust base for different enterprises and promoting cross-domain collaborative manufacturing \cite{hazra2021comprehensive,huo2022comprehensive}. However, blockchain is a deterministic and closed system that cannot actively access external data. To resolve this limitation, blockchain oracle is proposed as the bridge  between blockchain and external data \cite{pasdar2022connect}.

However, the features of IIoT, such as rapid data changes, network, and device heterogeneity, pose challenges to the quality of service of the oracle. For example, when the smart contract needs to obtain real-time sensor data in IIoT, due to the existence of the above problems, the data obtained by different nodes may vary greatly, which brings difficulties to the data aggregation process and affects the service quality of the oracle. To improve the service quality of the oracle, some research schemes apply the reputation mechanism to the oracle and improve the quality and credibility of the obtained data by encouraging and selecting high-quality nodes. ChainLink \cite{chainlink} is one of the classic representatives. It evaluates the service quality and credibility of the oracle node by establishing a reputation mechanism. The service provider with a higher reputation will receive considerable benefits, thereby improving the service quality of the oracle. Taghavi et al. \cite{taghavi2023reinforcement} used reinforcement learning to score nodes and select high-quality nodes to complete tasks to improve service quality. 
However, since the non-anonymous node selection process in the above schemes is open and transparent on the blockchain, it may be subject to malicious attacks (such as Sybil attack\cite{douceur2002sybil}, target attack\cite{sun2020voting}, etc.) and return incorrect data, endangering the security of the system.
Therefore, to ensure security in the node selection process, a node selection algorithm based on VRF is proposed to ensure anonymity security in the node selection process with its unpredictable but verifiable characteristics \cite{dos, lin2022novel}. However, the ensuing problem is a completely random node selection scheme, which is difficult to guarantee the quality of service of the oracle. 
Therefore, a key issue is how to design an oracle scheme to improve the quality of service of the selected nodes while ensuring the anonymous security of the node selection process, and improving the data consistency of the oracle in complex scenarios such as IIoT, thereby improving its quality of service. Although extensive research has been conducted on blockchain oracles, no study has been conducted to completely solve the above problems.

Considering the above issues, this paper proposes a blockchain oracle scheme for IIoT that considers security, credibility, and quality of service. Firstly, we design a node selection algorithm that combines a reputation mechanism and a VRF to select nodes with a high reputation secretly. Then, we propose a data filtering algorithm based on a sliding window to improve the consistency and efficiency of data collection and improve the service quality of the system. We believe that the scheme is not only suitable for the IIoT but also can be used as a general framework for the oracle, providing new ideas for improving the service quality of the blockchain oracle.

The main contributions of this paper are summarized as follows.

    \begin{itemize}

    \item We design a node selection algorithm that combines reputation mechanisms and VRF to address the tradeoff between node selection security and service quality. This algorithm ensures the anonymity and randomness of node selection while improving the service quality of selected oracle nodes, thus enhancing the security and service quality of the oracle network.

    \item To address the issue of low-quality real-time data obtained from heterogeneous oracle nodes, we design a data filtering algorithm based on sliding windows. It reduces the time difference and variance of the data obtained between nodes, incentivizes oracle nodes to obtain data from the data source as quickly as possible, and improves the consistency and efficiency of data acquisition.
    
    \item The security of the proposed scheme is verified through security analysis, and simulation experiments demonstrate that our scheme can effectively improve the security and service quality of the oracle network. When there is a 10\% malicious node occupancy rate in the network, our proposed scheme increases the data accuracy by approximately 4\% and reduces the average data variance by about 45\%, compared to conventional schemes.
    \end{itemize}
    
    The remaining parts of this paper are organized as follows. Chapter \ref{related_work} presents some related work that is relevant to our research. Chapter \ref{preparation} introduces the necessary background knowledge, such as oracle and verifiable random functions. Chapter \ref{work} describes the oracle scheme proposed in this paper. Chapter \ref{security_analysis} provides a security analysis of the proposed scheme. In Chapter \ref{result}, experimental results demonstrate the significant advantages of our proposed oracle scheme in terms of service quality, security, and robustness. Finally, Chapter \ref{conclusion} concludes the paper.

\section{RELATED WORK}
\label{related_work}

The following will introduce the existing decentralized oracle schemes.

\subsection{Voting-based Oracle}
Augur \cite{peterson2015augur} is the first proposed decentralized oracle prediction market platform for Ethereum. It distributes data request events to all participants in the market, and most people agree that the answer is rewarded, otherwise, it is punished. Astraea \cite{adler2018astraea} divides participants into three categories: submitters, voters, or verifiers. Under the assumption of rational people, the honest behavior of each participant is the Nash equilibrium solution of the system. On this basis, Cai et al. \cite{cai2022truthful} jointly determined the reward of voters through a lightweight scoring mechanism and a nonlinear voting weight scaling mechanism to avoid herding \cite{galariotis2015herding}. However, the above schemes have the problem of the need to determine results and the long voting period, which is not suitable for returning data sources with frequent result changes, such as real-time sensor data in IIoT.

\subsection{Hardware-based Oracle}
Trusted Execution Environment (TEE) can protect the core code of the program from interference by other malicious programs, while avoiding data leakage \cite{jauernig2020trusted}. Zhang et al. \cite{zhang2016town} designed the Town crier oracle scheme, which performs authentication and data request processing in TEE to ensure the security and credibility of acquired data. Based on the Town crier, Woo et al. \cite{woo2020distributed} improved the availability of oracle networks in the event of a single point of failure or the emergence of malicious nodes through the Byzantine Fault Tolerant (BFT) consensus mechanism. Liu et al. \cite{liu2022extending} provided trusted vaccine anticounterfeiting data for the blockchain through a low-cost Microcontroller Unit (MCU) that supports TEE. However, in the actual industrial IoT scenario, heterogeneous devices are difficult to guarantee support for TEE, and there are already many studies on TEE attacks\cite{jauernig2020trusted}.

\subsection{Cryptography-based Oracle}
The oracle schemes based on cryptography can be roughly divided into two categories: Transport Layer Security (TLS) and threshold signature. The representative scheme of TLS is DECO\cite{zhang2020deco}, which is based on zero-knowledge proof and TLS, allowing the oracle node to prove that the data comes from the specified Web data source without relying on trusted hardware. The threshold signature scheme can ensure that the data is not disclosed before aggregation is complete. Only by collecting the signature fragments that meet the threshold can an effective signature be generated to prevent the "freeloading" problem\cite{chainlink}. DOS Network\cite{dos} and Lin et al.\cite{lin2022novel}  first select a set of nodes through a VRF to obtain data, and then sign the data with a threshold signature algorithm. Similarly, Manoj et al.\cite{manoj2023trusted} obtained trusted agricultural data for the blockchain using a threshold secret-sharing scheme. The problems with the above schemes are also obvious. TLS requires a trusted certificate authority, which is contrary to the concept of blockchain decentralization. The low reliability of devices in the heterogeneous Internet of Things reduces the availability of threshold signatures, and it is difficult for threshold signatures to reach a consensus on frequently changing data \cite{vari2021issues}.

\subsection{Reputation-based Oracle}
Although the first three methods have been extensively studied, they are difficult to deal with rapidly changing data and heterogeneous network devices in the IIoT. The reputation mechanism selects and motivates nodes with high reputations to perform tasks by evaluating nodes. Its fewer constraints make it easier to adapt to the complex and changing environment of the IIoT. ChainLink \cite{chainlink} evaluates the service capability and credibility of the oracle node by establishing a reputation mechanism. Service providers with higher reputations receive significant benefits, ensuring high availability and data authenticity. For the lazy behavior of nodes, Du et al. \cite{du2022novel}, based on an auction mechanism, improved the enthusiasm of individual feedback data through more reasonable incentives to meet the specific delay constraints of smart contracts. Taghavi et al. \cite{taghavi2023reinforcement} used reinforcement learning to score nodes and select high-quality nodes to complete tasks to improve service quality. In addition, there has been a lot of research on reputation mechanisms in blockchain. For example, Beh-Raft-Chain \cite{wang2020beh} and RepuCoin \cite{yu2019repucoin} improve the security and efficiency of blockchain by considering the reputation of nodes when selecting consensus nodes. However, the non-anonymous node selection process, whether based on reputation-weighted or random, can easily lead to node collusion or becoming the target of malicious attacks. Moreover, the Matthew effect\cite{merton1968matthew} makes it difficult for new nodes to compete with the original nodes, resulting in fewer new nodes willing to join, which is not conducive to the improvement of oracle service quality.
Goel et al. \cite{goel2021infochain} proved that random node selection can effectively prevent collusion among nodes, under the reasonable assumption that miners always implement smart contracts honestly (otherwise they risk not receiving block rewards). Although the VRF-based node selection algorithm \cite{dos,lin2022novel} can reduce the risk of nodes being predicted or exposed to targeted attacks in the selection process, the quality of the selected nodes cannot be guaranteed.

\begin{table}[h] 
\caption{Comparison of relevant literature.}
\label{table:related_works}
\resizebox{\linewidth}{!}{
\begin{tabular}{|l|c|c|}
\hline
 Literature & Anonymous security & Quality of service \\ \hline
   \cite{peterson2015augur,adler2018astraea,zhang2016town,liu2022extending,zhang2020deco,goel2021infochain,manoj2023trusted,cai2022truthful} & \ding{55} & \ding{55} \\ \hline
    \cite{woo2020distributed,du2022novel,chainlink,taghavi2023reinforcement,wang2020beh,yu2019repucoin}& \ding{55} & \ding{51} \\ \hline
    
    \cite{dos,lin2022novel} & \ding{51} & \ding{55} \\ \hline
    
    Ours & \ding{51} & \ding{51} \\ \hline
\end{tabular}
}
\end{table}

Although blockchain oracles have been extensively studied, no study has addressed the tradeoff between security and service quality in oracle node selection. Table \ref{table:related_works} analyzes the existing research from the perspective of anonymous security and quality of service. In addition, few studies have explored the relationship between the heterogeneity of oracle nodes and networks and data quality. Therefore, this study aims to ensure the anonymity and randomness of the oracle node selection process while selecting high-quality nodes and to reduce the problem of low data quality caused by the heterogeneity of nodes and networks.

\section{PRELIMINARIES}
\label{preparation}

\subsection{Oracle}

\begin{figure}[!ht]
    \centering
    \includegraphics[width=3in]{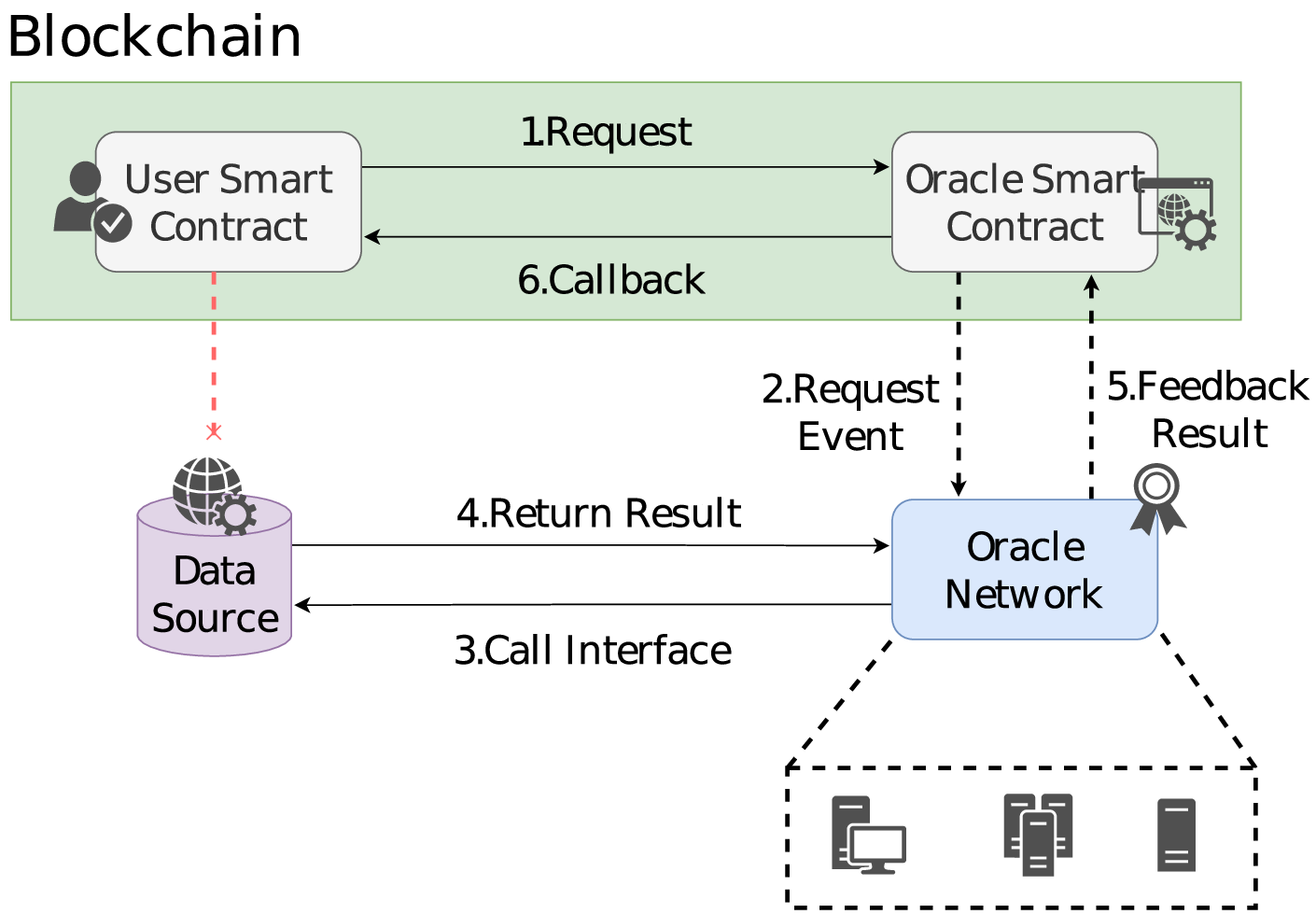}
    \caption{Overview of oracle system.}
    \label{fig:Oracle}
\end{figure}

Smart contracts and blockchains are similar to a closed system that cannot access external information, which means that interactions are limited to the data available on the blockchain. This is an open practical problem, known as the oracle problem, which is defined as how to transmit real-world data to the blockchain \cite{pasdar2022connect}.

The main steps of the solution are illustrated in Fig. \ref{fig:Oracle}: 1) The user contract initiates an on-chain request by calling the function of the oracle contract, and then the oracle contract records the event for the oracle according to the user contract's request; 2) After the request event is recorded in the blockchain event log through the consensus, any node in the oracle network listens for request events from the oracle contract; 3) The oracle node requests data or executes operations from third-party services based on the request event; 4) All nodes in the oracle network that have completed the request processing aggregate the obtained results through a specified aggregation mechanism to obtain a final feedback result; 5) The oracle network uploads the final feedback result to the oracle contract; 6) After obtaining the feedback result, the oracle contract returns it to the user contract in the form of a callback.

\subsection{Verifiable Random Function}

Verifiable Random Function (VRF)\cite{micali1999verifiable} is a type of pseudorandom function that can generate corresponding pseudorandom numbers and non-interactive proofs based on data input, and anyone can verify the correctness of the random numbers through the proof. VRF can be represented as a triad of algorithms with polynomial-time complexity $(VRF_{Setup}, VRF_{Generate}, VRF_{Verify})$. $(sk,pk) \gets VRF_{Setup} ()$ generates a private key $sk$ and its corresponding verification public key $pk=g^{sk}$, where $g$ is a generator element of the cyclic group; $(y,\pi) \gets VRF_{Generate} (sk,x)$ means using the private key $sk$ to encrypt any message $x$ to obtain a pseudorandom string $y$ and its non-interactive proof $\pi$; $0/1\gets VRF_{Verify} (pk,x,y,\pi)$ means using the verification public key $pk$ to verify the effectiveness of the proof $\pi$, that is, whether $y$ is the pseudorandom string obtained by encrypting the message $x$ using $sk$.

For any given key pair $(sk,pk)$ and input $x$, VRF guarantees that there is no other valid output $y^{'}$ and proof $\pi^{'}$ that can make $VRF_{Verify}$ valid. In addition, VRF also has unpredictability, that is, its output is unknown to anyone before the private key $sk$ is made public. By combining this feature with existing distributed threshold signature schemes, a decentralized random number generation algorithm can be constructed, providing a reliable source of random numbers for probabilistic dependent distributed systems\cite{dos}. Furthermore, using VRF to select consensus nodes in the blockchain can effectively reduce targeted attacks after the node identity is revealed\cite{chen2022blockchain}.

\section{SYSTEM DESIGN}
\label{work}

    \begin{figure*}[!ht]
        \centering
        \includegraphics[width=5in]{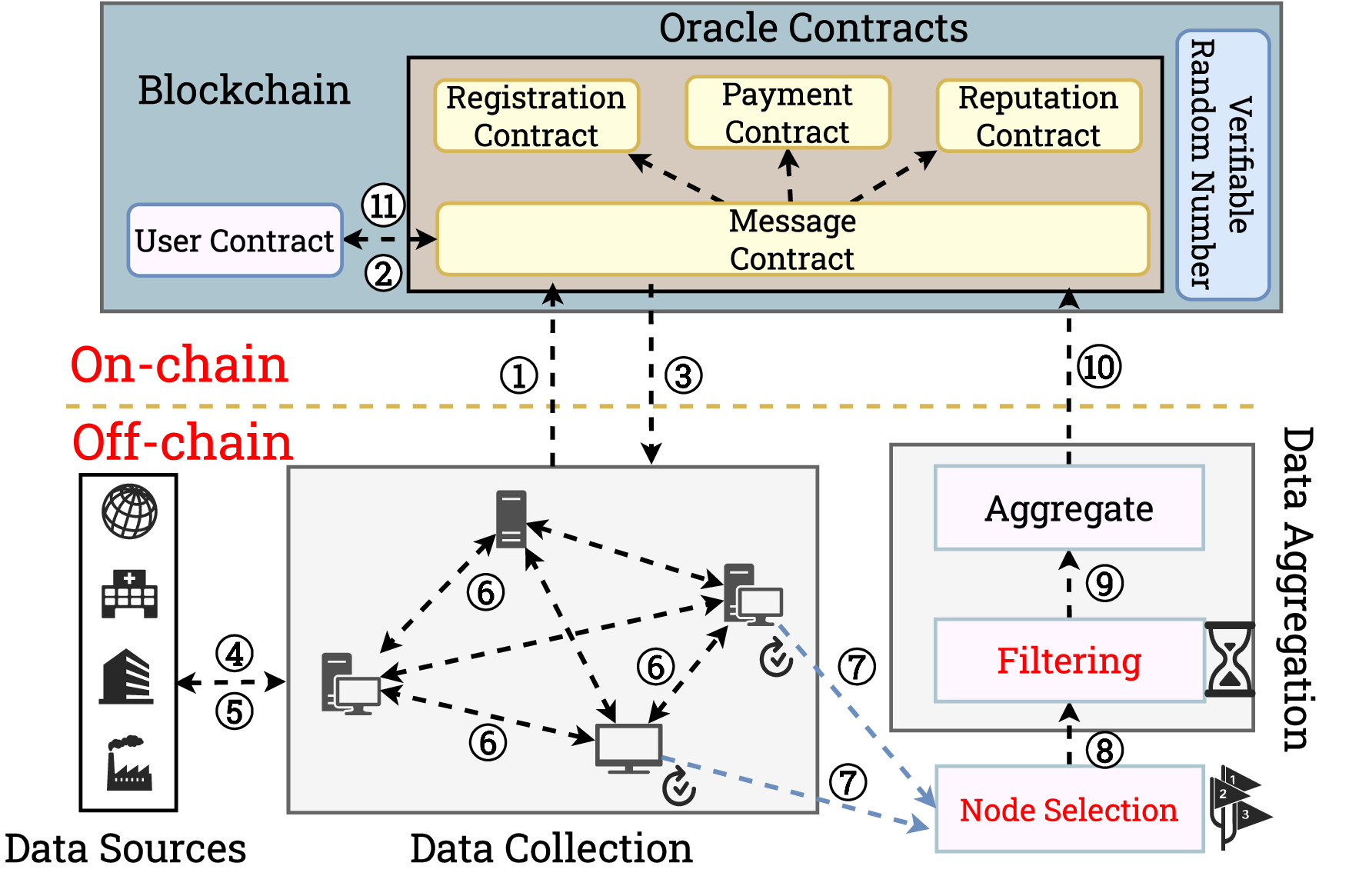}
        \caption{Overview of our oracle scheme.}
        \label{fig:Overview}
    \end{figure*}

In this section, we provide a detailed description of the proposed oracle scheme. The goal of this scheme is to provide trusted, high-quality off-chain data for the IIoT blockchain and maintain the trust foundation for cross-domain collaboration among IIoT participants.

\subsection{System Structure}
As shown in Fig.\ref{fig:Overview}, the proposed oracle solution in this paper consists of four parts, on-chain oracle contract, data aggregation module, data collection module, and node selection algorithm. The on-chain oracle contract is responsible for processing requests from user contracts and recording corresponding request events. The aggregation module can process the data returned by the oracle according to user-defined aggregation rules to produce a final result. The data collection module is responsible for collecting and verifying the results submitted by the oracle. The node selection algorithm is responsible for secretly selecting a subset of available oracle nodes from the IIoT network to process specific data request events.

\subsubsection{Oracle Contract}
The on-chain oracle contract is the communication medium between user contracts and the oracle network, which provides message delivery and management of oracle nodes. The oracle contract in this paper mainly consists of the following four sub-contracts:

    \begin{itemize}
        \item \textbf{Registration Contract}: The blockchain node can register to join the oracle network only after pledging a certain amount of token assets to the contract (because the source of virtual assets depends on the specific incentive mechanism of the blockchain system, this article does not discuss its source, only assuming that participants hold a certain amount of token assets when registering). After successfully joining, the public key corresponding to the node account will be recorded in the registration contract, and other participants can query the public key from the contract to verify the identity of the node publishing data. In addition, when a node commits malicious acts, the registration contract can confiscate its deposit and use it to reward other honest nodes.
        
        \item \textbf{Message Contract}: When a user contract needs to request data from an oracle, it needs to call a function through the message contract. Once the call is successful, the message contract will generate and record the corresponding data request event. The off-chain oracle nodes will continuously listen for request events from the message contract, request data from the specified data source, and feedback on the final result of the message contract. After receiving the result from the oracle, the message contract will return the result through a callback function specified by the user contract, enabling the user contract to perform subsequent computations after successfully obtaining the data.
        
        \item \textbf{Payment Contract}: The oracle service incurs certain costs while it is running. For example, when the message contract calls back the user contract, it needs to pay a certain number of transaction fees to the blockchain. At the same time, the off-chain oracle network also incurs corresponding costs when providing services to user contracts. Therefore, the user contract should pay a certain fee to the oracle service as a reward. The provider of the user contract must first deposit a certain amount of funds into the payment contract to obtain permission to use the message contract to access the oracle service. After each request event is processed and the correct data is successfully returned, the payment contract will pay the corresponding service fee to the oracle node that provided the accurate data.
        
        \item \textbf{Reputation Contract}: The reputation of an oracle node is calculated by considering its historical behavioral data, including the average response time to events, the accuracy of feedback data, and the total number of requests processed. The reputation contract will record the above data and calculate the reputation value of the node according to specific calculation rules. Fast and accurate data feedback will increase the reputation value of the oracle node, while the opposite may cause its reputation to decline. A higher reputation in the node selection module can help the oracle node increase its probability of being selected. Considering both oracle performance and data quality, the reputation value $R_i$ of any oracle node $O_i$ is represented as: 
        
        \begin{equation}
            \begin{aligned}
                R_i = (log S_i) ( \frac{\alpha}{T_i} + (1-\alpha) \times A_i ), \alpha \in (0,1)
            \end{aligned}
        \end{equation}
        
        Here, $T_i$, $A_i$, and $S_i$ represent the average response time, response data accuracy, and total service times of node $O_i$ ($1 \leq S_i$), respectively. 
        $\alpha$ is a system hyperparameter used to adjust the preference of the node selection algorithm for nodes. Its increase and decrease represent the weight of improved node performance (faster response time) and data quality (higher correctness) in the reputation value, respectively.
        In addition, $log S_i$ can significantly improve the reputation value of nodes in the early stages of joining the network, while preventing unlimited reputation growth with increasing service times $S_i$ and avoiding the Matthew effect. When the reputation drops to a certain value, the assets pledged by the node in the registration contract are confiscated to encourage oracle nodes to provide higher-quality services.
        
    \end{itemize}

\subsubsection{Node Selection Module}
The node selection algorithm can implicitly select a subset of registered oracle nodes to process specific data request events on the blockchain. The algorithm is fully decentralized and oracle nodes continuously listen to data request events on the chain and calculate their priority in the request event based on the current verifiable random number. Higher priority means that the node's response to the request event is given priority in the subsequent sorting process. Oracle nodes can decide whether to provide services based on their priority in the event. In addition, due to the inherited properties of the verifiable random function, each node's priority is unpredictable but verifiable to other nodes. That is, before providing feedback, the outside world cannot predict its specific priority in a particular request event; after providing feedback, the outside world can verify whether the published priority is true. These features maintain the anonymity of participating nodes and effectively prevent targeted attacks on the operation of the oracle service. We will discuss the random node selection algorithm and priority calculation method in detail in \ref{node_select}.

\subsubsection{Data Collection Module}
The data collection module runs on each oracle node, continuously collecting and verifying the feedback results broadcasted by other nodes for data request events (including timestamp, proof of data source for data authenticity, submitter's identity, and their priority as the current request event handler), to obtain a qualified result set. Its main function is to prevent the "freeloading" problem. In the solution to the "freeloading" problem, this paper adopts the two-phase threshold signature scheme based on ChainLink. The biggest feature of this scheme is that all feedback results are initially submitted in ciphertext form. Only when the feedback results of the threshold $t$ oracle nodes for the same request event are aggregated and signed into an effective signature, all participating nodes will publicly reveal their private keys to decrypt and disclose the feedback results. This mechanism effectively prevents cheating behavior by nodes trying to reduce data acquisition costs by copying feedback data from other nodes.

\subsubsection{Data Aggregation Module}
The data aggregation module aggregates the feedback results from the oracle nodes according to the rules defined by the user and returns them to the oracle contract. Different aggregation strategies need to be used for different data and business needs. For example, for data with frequent fluctuations in a short period, the median or mean is usually used as the final result. For data with longer change cycles, it is usually accepted after undergoing consistency checks. This module may be implemented on-chain as a smart contract or off-chain using consensus protocols, depending on the specific application scenario. Before aggregation, two steps need to be completed: 1) Sorting all feedback results according to the priority of the corresponding nodes and selecting the results from the top t nodes with the highest priority; 2) Using a sliding time window-based data filtering algorithm to filter out data with a more concentrated time distribution from the data selected in the previous step, to improve the quality of subsequent aggregation results. 
We will discuss the data filtering algorithm in detail in \ref{filtering}.

\subsection{System Flow}

The oracle workflow is as shown in Fig.\ref{fig:Overview}:

\begin{enumerate}[label=\circled{\arabic*}]
    \item The off-chain node $O_i$ registers with the oracle contract by depositing a certain amount of tokens and uploading its public key $pk_i$ as a unique identifier. Upon successful registration, the reputation contract assigns an initial reputation value $R_i$ to the node, indicating its likelihood of being selected in the subsequent node selection process.
    
    \item A user contract can send a data request $q$ to the oracle contract after paying a service fee for a reward in the payment contract. The oracle contract records the data request event $E=(q,d,f,t,w)$ on the blockchain. Here, $q$ can serve as the unique identifier of $E$, $d$ represents the set of data sources specified by the user contract, $f$ represents the reward that the user contract needs to pay to all oracle nodes upon a successful request, $t$ represents the minimum number of nodes required to process the request, and $w$ represents the maximum width of the time window (i.e., the maximum time difference between feedback result timestamps).
    
    \item Upon listening about a new request event $E$ on the blockchain, any oracle node $O_i$ calculates its priority $L_{q,i}$ based on its private key $sk_i$, reputation value $R_i$, the latest verifiable random number $\xi_r$ known in the current blockchain, and the unique identifier $q$ contained in the request event. It then decides whether to participate in the service for the request event $E$. Note that the node selection process is not mandatory, and nodes can keep their priorities confidential without any impact. If the priority $L_{q,i}$ of a node is relatively high, i.e., there is a greater probability that it will provide feedback data for the request and receive corresponding rewards, the node will be more inclined to participate in the service.
    
    \item Assuming that node $O_i$ decides to respond to the request event $E$, it must select a specific data source from set $d$ based on the result of taking the modulus of its priority $L_{q,i}$ with the size $|d|$ of the data source set, and request data. Based on the verifiability of the proposed node priority, other nodes can confirm whether $O_i$ honestly obtained the data from the specified data source by verifying $O_i$'s public $L_{q,i}$ and data source proof in the feedback result.
    
    \item The data source processes the request from node $O_i$ and returns real-time data $x_{q,i}$ as the result. According to the method proposed in \cite{zhang2020deco}, $O_i$ needs to generate a proof $p_{q,i}$ for the communication process, proving that $x_{q,i}$ is obtained from the specified data source at time $ts_{q,i}$.
    
    \item Node $O_i$ broadcasts the result $I_{q,i} = (m_{q,i},pk_{q,i},\eta_{q,i},p_{q,i},ts_{q,i},L_{q,i})$ to the oracle network, where $(sk_{q,i},pk_{q,i})$ is the temporary key pair generated by $O_i$ for this submission. $\eta_{q,i}$ represents the partial signature of $q$ obtained by threshold signing using $sk_{q,i}$. $m_{q,i}$ represents the ciphertext obtained by partially encrypting $x_{q,i}$ using $sk_{q,i}$.

    \item Nodes participating in the service will continue to collect the results broadcast by other nodes and verify $p_{q,i}$ until $t$ $\eta_{q,i}$ from different nodes are aggregated by any node at least. If any node $O_j$ collects feedback results that meet the threshold requirement of $t$, it can compute a group signature $\eta_q$ and make it public. Once node $O_i$ detects a valid group signature $\eta_q$, it will make its temporary private key $sk_{q,i}$ public so that other nodes can decrypt $m_{q,i}$ to obtain $x_{q,i}$. After decrypting all feedback results $I_{q,i}$, node $O_j$ sends the corresponding $K_{q,i}=(x_{q,i},p_{q,i},ts_{q,i},L_{q,i})$ to the node selection module.
    
    \item The node selection module receives results from several data collection nodes and removes duplicates, sorts them by priority $L_{q,i}$, and obtains feedback results from the $t$ nodes with the highest priority. Thus, the selection process of the oracle nodes can be considered complete.
    
    \item Based on a sliding time window data filtering algorithm, the data with timestamp $ts_{q,i}$ selected by the $t$ nodes can be further filtered to obtain the final data $K_q$ that can be aggregated. After executing the aggregation logic on $K_q$, the data aggregation module writes the aggregation results and the information of the selected oracle nodes to the message contract.
    
    \item After receiving feedback from the data aggregation module, the message contract updates the reputation contract by updating the average response time, response data accuracy, and total service time information for the oracle nodes in that service. If a node's data feedback is filtered out by the data filtering algorithm or judged to be an outlier during the aggregation process, it is considered to be incorrect data, resulting in a decrease in response data accuracy. Finally, the payment contract pays a specified reward to oracle nodes that return correct data.
    
    \item Finally, the message contract feeds the aggregation results back to the user contract by callback, so that the user contract can complete the subsequent calculation process.
    
\end{enumerate}

\section{KEY ENABLING TECHNIQUES}

\subsection{Node Selection}
\label{node_select}

    After the blockchain nodes register as oracle nodes with the registration contract, they can become potential candidates for processing subsequent data requests. As mentioned earlier, the node selection algorithm in this paper is mainly implemented based on priority sorting, and each oracle node has a different priority for different oracle events. Therefore, the core of the node selection algorithm is how to design a random priority calculation method. It should take into account the credibility of the node in the calculation process, and allow nodes with higher credibility to have a certain probability of receiving a higher priority while maintaining randomness. Furthermore, the priority of a node is unknown before it is published, but can be quickly verified after it is published to ensure the anonymity of the node during the work process.
    
\begin{figure}[!ht]
            \centering
            \includegraphics[width=3in]{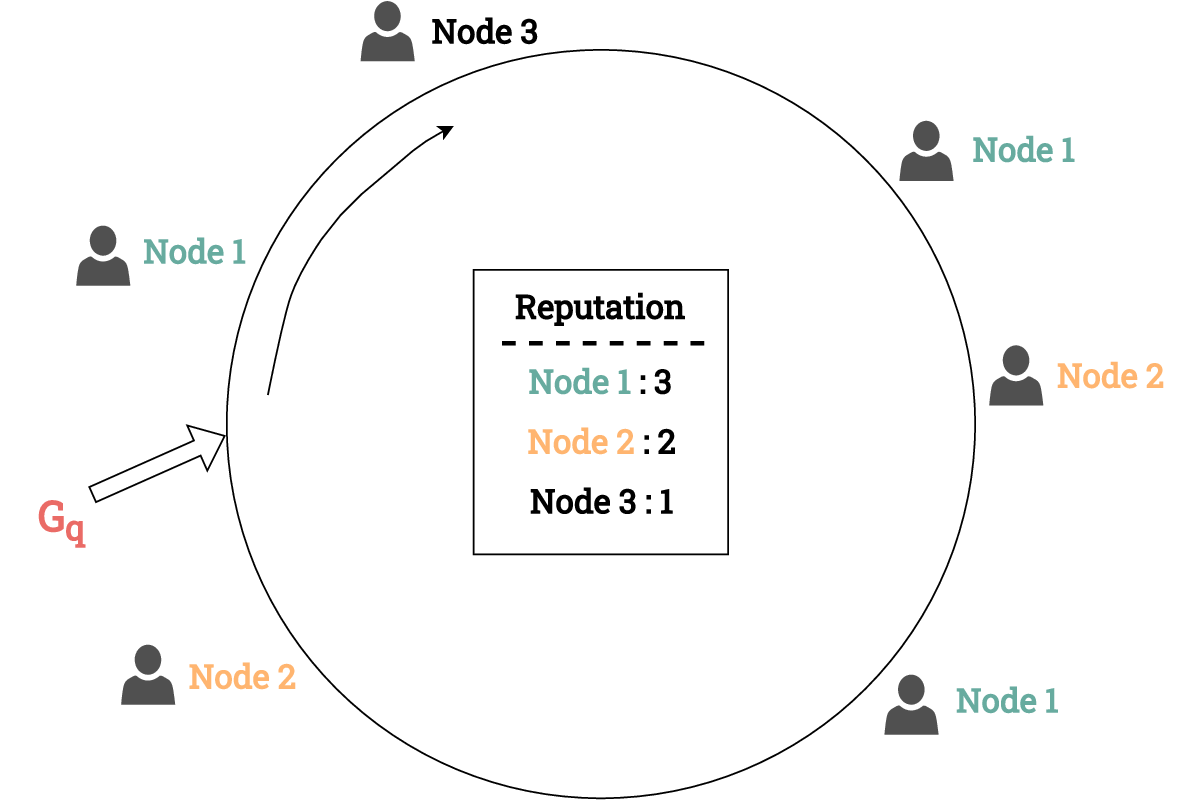}
            \caption{Diagram of oracle node selection algorithm.}
            \label{fig:hash}
        \end{figure}

    This paper's random node selection algorithm is based on the idea of the consistent hashing function \cite{karger1999web}. Given the hash function $H$, its range is organized into a virtual circle, called a hash ring, in a clockwise direction. As shown in Fig. \ref{fig:hash}, the hash function can map request events and all oracle nodes to different positions on the hash ring, and then oracle nodes determine their priority based on the distance between their position on the hash ring and the position of the request event on the hash ring. The closer a node's position is to the request event's clockwise distance, the higher its priority. In short, only the top $t$ nodes with the closest hash values need to be selected based on their distance on the hash ring.
    
    Given a hash function $H$, assuming the latest verifiable random number maintained in the current blockchain is $\xi_r$. When node $O_i$ receives a new data request event $E$ from the blockchain, it first hashes the unique identifier $q$ and the random number $\xi_r$ of $E$ to obtain the corresponding position of $E$ on the hash ring:
    
    \begin{equation}
        G_q = H(q \parallel \xi_r)
    \end{equation}
    
    Since both $q$ and $\xi_r$ are public, all participants can compute $G_q$. As reputation needs to be taken into account when randomly selecting nodes, in our scheme, the number of positions a node occupies on the hash ring is determined based on its reputation. For example, when the reputation of $O_i$ is $R_i=10$, it will be mapped to 10 different positions on the hash ring, and it only needs to select the position with the smallest clockwise distance to $G_q$ to calculate the priority for this request. The higher the reputation of a node, the more positions it occupies on the hash ring, and the greater the probability of obtaining a high priority. When there are more oracle nodes, the probability of selecting high-reputation nodes can be increased while ensuring randomness. When $O_i$ needs to calculate its position on the hash ring, it signs $q$ and $\xi_r$ with the signature function $Sign$ and its private key $sk_i$ to obtain:
    
    \begin{equation}
        g_{q,i} = Sign(q \parallel \xi_r \parallel sk_i)
    \end{equation}
    
    Since $\xi_r$ cannot be predicted in advance until the moment it is generated \cite{dos}, $O_i$ cannot be predicted to compute the high priority $g_{q,i}$. Then, node $O_i$ needs to query its reputation value $R_i$ from the reputation contract on the chain and use $g_{q,i}$ as the generator to compute the set of all positions mapped to the hash ring for this request:
    
    \begin{equation}
        Y_{q,i} = \{H(g_{q,i}^k) \mid 1 \leq k \leq \left \lceil R_i \right \rceil \}
    \end{equation}
    
    Finally, calculate the shortest distance from all positions in $Y_{q,i}$ to $G_q$. The shorter the distance, the higher the priority of the node in processing the request event $E$. Therefore, the priority of node $O_i$ can be expressed as: 
    
    \begin{equation}
       L_{q,i} = \frac{1}{min(\mid G_q - y \mid)}, y \in Y_{q,i}
    \end{equation}

\subsection{Data Filtering}
\label{filtering}

    Due to the heterogeneity of nodes and the differences in geographical location in the IIoT, the delay in obtaining data varies among different nodes. In scenarios where data frequently changes, data consistency may be poor. The data filtering algorithm in this paper focuses on filtering the data before aggregation. It filters to obtain a portion of the data with a more concentrated temporal distribution according to the width of the time window specified by the oracle contract. Assuming that the data aggregation module selects $\{K_{q,i} \mid 0 \leq i < t\}$ and corresponding timestamps $\{ts_{q,i} \mid 0 \leq i < t \}$ based on the priority of the feedback results submitted by nodes, and the time window width limit is $w$. The filtered data set $K_q$ should meet the following three requirements: 1) The range of time stamps of the data is no greater than $w$, to avoid large differences between two feedback results; 2) The number of data contained in the time window should be as many as possible; 3) When the amount of data is the same, the variance of the data timestamps should be as small as possible, making the overall data acquisition time closer to the expected value, i.e., the distribution is more concentrated.
    
\begin{figure}[!ht]
        \centering
        \includegraphics[width=3in]{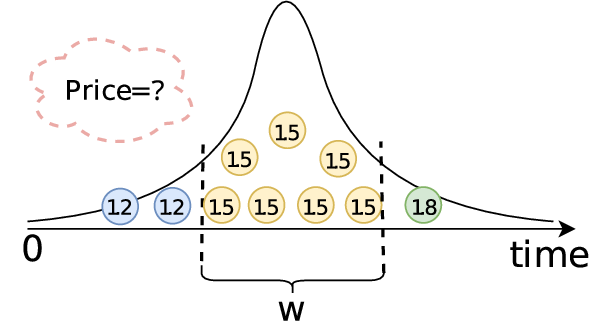}
        \caption{Diagram of data filtering algorithm.}
        \label{fig:data_verify}
    \end{figure}

    As shown in Fig. \ref{fig:data_verify}, the selected oracle nodes get different results from the data source at different times. We use a time window of width $w$ to move from left to right, count the variance of the data timestamp in all time intervals, and filter out the data in the time interval with the least variance. The specific implementation of the reference algorithm is \ref{algorithm:data_verify}.

    \begin{algorithm}[]
        \caption{Data filtering algorithm based on sliding window}
        \label{algorithm:data_verify}
        \LinesNumbered
        \KwIn {Feedback results of nodes $K_{q,0},…,K_{q,t-1}$ , Timestamp of data $ts_{q,0},…,ts_{q,t-1}$, Time window width $w$}
        \KwOut {Contains the subset of results with the most data and the smallest time variance $K_q$}
        \BlankLine
        $l,r = 0,0$\;
        $maxnum, minvar = 1,0$\;
        $K_q = \{ K_{q,0} \}$\;
        \While{$r \leq t-1$}{
            \eIf{$ ts_{q,r}- ts_{q,l} \leq w$} {
                $num = r-l+1$\; 
                $avg = \sum_{i=l}^{r}ts_{q,i}/num$\;
                $var = \sum_{i=l}^{r}(ts_{q,i} - avg)^2 / num$\;
                \If{$maxnum \leq num \ or \ (num == maxnum \ and \ var < minvar)$}{
                    $maxnum,minvar = num,var$\;
                    $K_q = \{K_{q,i} \mid l \leq i \leq r \}$\;
                }
                $r = r+1$\;
            }{
                $l = l+1$\;
            }
        }
        \textbf{return} $K_q$;
    \end{algorithm}

    The data filtering algorithm can motivate the oracle nodes to obtain data from the data source as quickly as possible. Assume that the delay of all nodes to the data source follows a normal distribution $N(\mu, \sigma^2)$ with a mean of $\mu$ and a standard deviation of $\sigma$ \cite{fraleigh2003provisioning}. That is, the delay of a small number of nodes to the data source may be lower or higher, but according to the concentration of the normal distribution function, the delay of the majority of nodes is concentrated in the interval closer to the average delay $\mu$. Therefore, to improve the probability of getting rewards, i.e., the probability of getting feedback data selected by the sliding time window, all nodes should get data from the data source and get feedback as soon as possible. Suppose a node deliberately delays processing the request, then even if it has a higher priority in this service, its feedback result is likely to be excluded from the sliding time window due to a larger time variance. In addition, if $w$ is set too small, it is difficult to ensure the security and credibility of the system. At this point, we can set a lower threshold $\zeta$ to reduce this risk. Tasks with aggregation numbers below the threshold will fail and increase $w$ by half before restarting.

    The above algorithm is relatively easy to implement when using the on-chain aggregation method, only the corresponding calculation process needs to be implemented in the smart contract. However, this method has obvious efficiency problems. The aggregation mechanism is mainly divided into on-chain and off-chain methods. In the on-chain aggregation method, the nodes write the results directly into the blockchain through the oracle contract after obtaining the results. When the aggregation condition is met, the oracle contract aggregates all the results according to the aggregation rules. When the off-chain aggregation method is used, the final result is obtained and fed back to the oracle contract by the oracle nodes through a special off-chain protocol. The difference between on-chain and off-chain aggregation is that the former is easy to implement, but will incur large on-chain computation and storage costs, while the latter is the opposite. If the number of participating oracle nodes is large, the submission process of on-chain aggregation may generate many transactions, which may consume a lot of time in the blockchain consensus phase and significantly affect the real-time performance of the oracle. Therefore, the off-chain aggregation method is more suitable for IIoT scenarios.
    
\begin{figure}[!ht]
        \centering
        \includegraphics[width=3in]{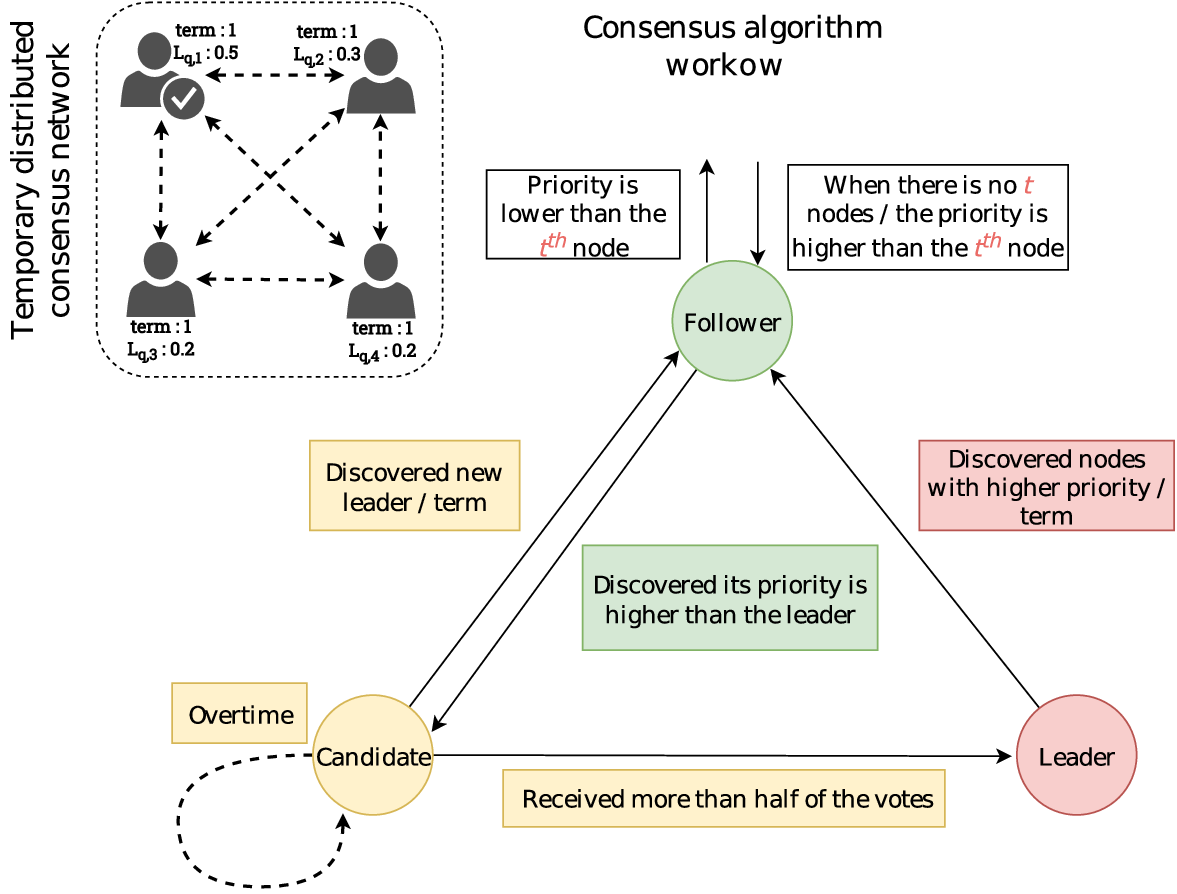}
        \caption{Diagram of distributed network construction.}
        \label{fig:raft}
    \end{figure}

    To run the above node selection algorithm and data filtering algorithm in a distributed manner during the off-chain aggregation process, this paper proposes an implementation solution based on the reputation distributed consensus algorithm Raft\cite{wang2020beh}, as shown in Fig. \ref{fig:raft}. First, it is necessary to form a temporary distributed consensus network by combining $t$ oracle nodes that participate in the data request event processing. When a node $O_i$ publishes its feedback result, the following situations may exist: 1) Other nodes have not returned results yet, then $O_i$ creates a new consensus network by broadcasting information and electing itself as the leader in the network; 2) There is already a consensus network for this service in the network, then $O_i$ joins as a following node. Suppose a consensus network already contains $t$ nodes, when $O_i$ tries to join the network, if its priority $L_{q,i}$ is higher than the lowest priority member $O_j$ in the current network, it can join successfully, and $O_j$ will be removed from the network, otherwise, the joining fails. Note that the leader node in the temporary consensus network is not always fixed, and the existing leader node may be removed due to attacks or low priority, which may lead to a new round of leader node elections.
    
    Secondly, all nodes in the temporary consensus network need to go through multiple rounds of consensus to complete the off-chain aggregation process: 1) In the first round, the consensus is reached with $t$ feedback results $\{ K_{q,i} \mid 0 \leq i < t \}$, and the consensus result is used as the input of Algorithm \ref{algorithm:data_verify} to obtain $K_q$; 2) In the second round, the consensus is reached with the $K_q$ obtained by each node, and the consensus result is used as the input of the aggregation strategy to obtain the aggregation result. Finally, the current leader node is responsible for submitting the aggregation result to the on-chain oracle contract, and all nodes in the temporary consensus network exit the off-chain aggregation process.

\section{SECURITY ANALYSIS}
\label{security_analysis}

    In this section, we present a security analysis of the proposed scheme.
    
    Since the selection of oracle nodes is publicly in the blockchain, attackers can launch various attacks, such as Sybil attack and Target attack, to gain profit. In addition, oracle nodes themselves may return incorrect data to gain profit. Undoubtedly, these attacks and the malicious behavior of nodes can threaten the security, reliability, and decentralization of oracles and blockchains. So, oracle systems should be designed to withstand attacks and malicious behavior of nodes. Therefore, we will analyze the security issues in the selection of oracle nodes from the perspective of cryptographic security and high economic security.

    \subsection{Sybil Attack}
    The Sybil attack refers to the practice of an organization or individual creating or using multiple accounts (false identities) in an attempt to manipulate or control a P2P network system \cite{douceur2002sybil}. In distributed oracle systems, Sybil attacks often involve a single person running multiple oracle nodes in an attempt to disrupt the normal operation of the service or gain more centralized power. In the proposed oracle service solution presented in this paper, nodes must pay a certain amount to participate in the service, making it difficult for malicious users to create unlimited accounts. Furthermore, even if a malicious user pays a high cost to obtain multiple accounts, the introduction of a random node selection algorithm and a reputation mechanism makes the probability that the majority of selected nodes are malicious in a single round very small, especially in IIoT environments where there are large numbers of nodes.

    \subsection{Targeted Attack}
    \label{target_accacks}

    The Target Attack is not a specific type of attack, but refers to the continuous attack launched by an attacker against a specific target\cite{sun2020voting}. Malicious users may attempt to disrupt the normal operation of the oracle service by launching a Target Attack on the nodes in the distributed oracle. In the data request phase of the proposed solution, the data request nodes are anonymous, and attackers are almost unable to predict all potential service nodes for a data request event. Even if an attacker successfully guesses a small number of nodes through monitoring network packets or other means and launches attacks on them to take them offline or release incorrect data, this situation will not significantly affect the normal operation of the proposed solution. The node selection algorithm in this paper does not directly select $t$ predetermined nodes, but is implemented by prioritizing nodes based on their priority in a certain request event. Even if a higher-priority node is offline due to an attack, it does not affect the system's ultimate selection of $t$ nodes. At the same time, the feedback data from the oracle nodes needs to be accompanied by corresponding TLS record proofs for other participants to verify the authenticity of the results. In the case where most participating parties are honest nodes and not colluding (as guaranteed by the random selection algorithm in probability), the data forged by the malicious nodes held by the attacker is almost impossible to pass the verification of other nodes.
    
    When an off-chain aggregation solution is used, nodes may no longer be anonymous because they need to reach a consensus, and they may be taken offline by malicious user attacks. In this case, the security guarantee of the off-chain aggregation solution comes from the consensus algorithm used at the underlying layer. The proposed solution uses the Raft \cite{ongaro2014search} consensus algorithm to ensure data consistency, so as long as no more than $t/3$ nodes are attacked, the aggregation process will not fail due to the withdrawal of these nodes. Replacing Raft with a Byzantine fault-tolerant algorithm can provide better security, but may reduce efficiency.

\section{PERFORMANCE EVALUATION}
\label{result}

    In this section, we verify the performance and security advantages of the proposed solution through simulations. Specifically, we first examine the performance of different solutions, then investigate the performance of the proposed solution under various proportions of malicious nodes, and finally analyze the reputation update and parameter effects on the solution's performance, demonstrating our research results.
    
    We simulated a smart contract system using the Golang language, which continuously publishes data requests and receives feedback data. We also use a distributed oracle to provide data acquisition services for the blockchain. To simulate the continuous data of real-time sensors, we design a data source that provides real-time feedback in the form of floating-point numbers. In an oracle network with 100 oracle nodes, we conduct 1000 oracle tasks. Baseline is an implementation based on schemes such as DOS Network \cite{dos,lin2022novel}. It uses VRF to randomly select nodes for tasks and has no data filtering module. We compared our scheme with the baseline and with schemes that remove some of the sub-algorithms (node selection with reputation weighting, and data filtering). The parameters of the experiment are shown in Table.\ref{table:para}.

    To ensure anonymous security in node selection, our scheme requires all participants to participate in the judgment at the beginning of the task, which increases the computational complexity. In the case of off-chain aggregation, compared with the baseline, our scheme mainly increases the priority of $O(Nlog(N))$ for node selection and the computational complexity of $O(t)$ for sliding time windows in data filtering. However, compared with the long communication time in heterogeneous networks, we believe that the computation time is negligible.

    \begin{table}[h]
    \caption{Experimental parameter settings.}
    \resizebox{\linewidth}{!}{
    \begin{tabular}{ccc}
    \hline
     Parameter & Meaning & Value \\ \hline
     $N$ & Total number of oracle nodes & 100 \\ 
     $\delta$ & Proportion of malicious nodes & 0.1 \\
     $t$ & Number of nodes selected per round & 10 \\ 
     $w$ & Length of time window & 1s \\ 
     $\alpha$ & The preference weight of reputation & 0.5 \\ 
     $\zeta$ & The minimum threshold number of aggregation & 1 \\
      \hline
    \end{tabular}
    }
    \label{table:para}
    \end{table}

    \subsection{Quality of Service}
    
    \begin{figure*}[!t]
    \centering
    \subfloat[Accuracy of data]{\includegraphics[width=0.33\linewidth]{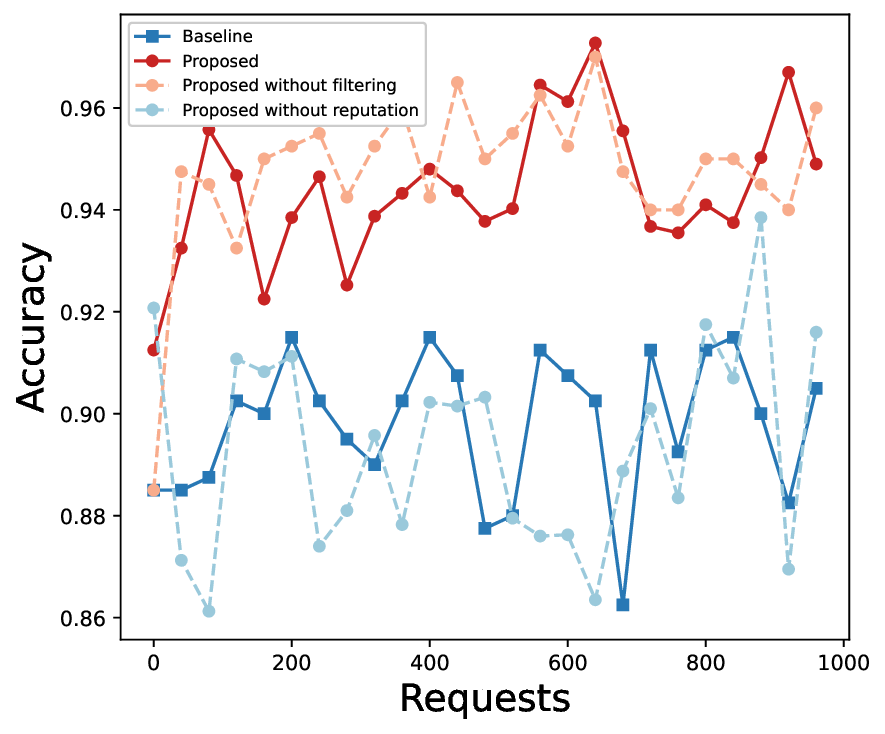}%
    \label{fig:ablation_acc}}
    \hfil
    \subfloat[Variance of data]{\includegraphics[width=0.33\linewidth]{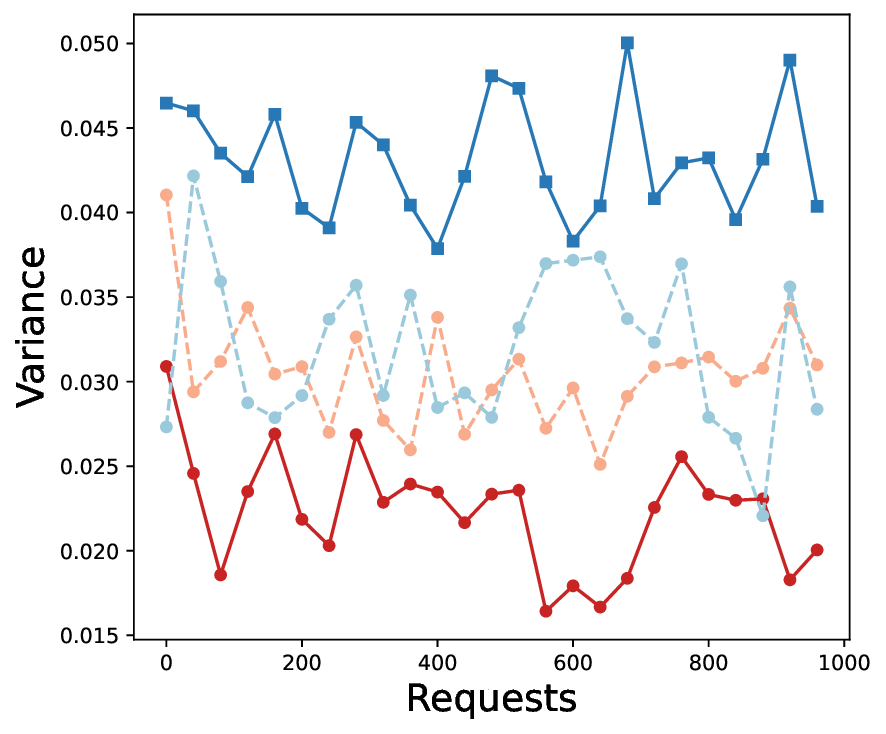}%
    \label{fig:ablation_var}}
    \hfil
    \subfloat[Time of obtaining data]{\includegraphics[width=0.33\linewidth]{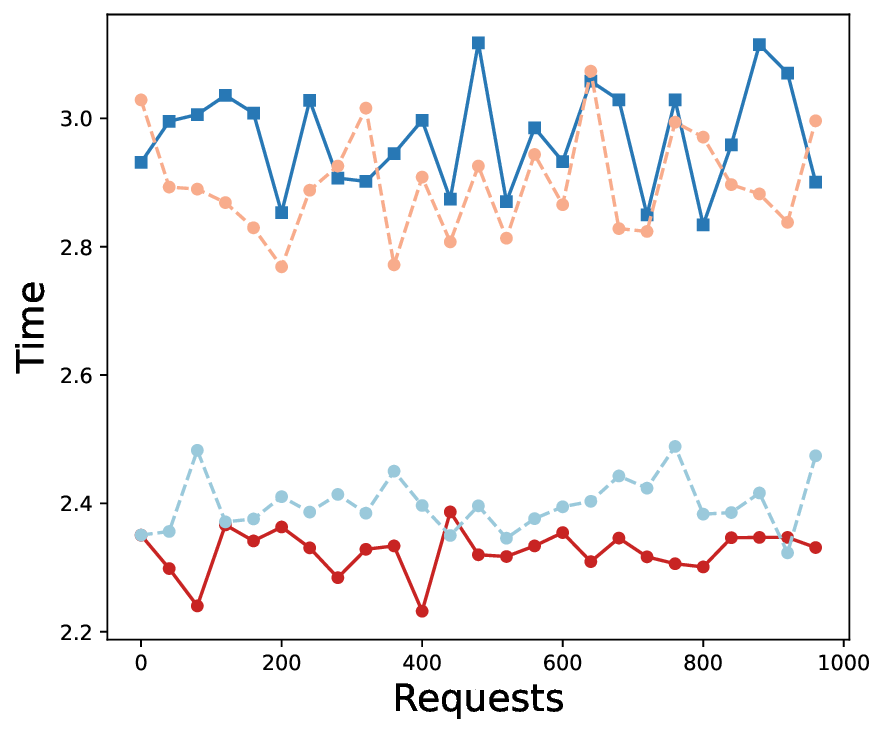}%
    \label{fig:ablation_time}}
    \caption{The accuracy, variance, and time of the data obtained by each scheme when there are 10\% of the malicious nodes.}
    \label{fig:ablation}
    \end{figure*}
    
    To demonstrate the performance advantage of our proposed approach over the baseline, we analyze the accuracy, variance, and time of the data obtained by our approach, the baseline, and the approach with some sub-algorithms removed (reputation-weighted node selection and data filtering), as shown in Fig. \ref{fig:ablation}. Variance is often used to measure the degree of deviation of a random variable from its expected value. Therefore, we use the variance of the returned data to reflect the consistency of the data, with lower variance indicating higher consistency.

    Fig. \ref{fig:ablation_acc} shows the accuracy of data obtained by different approaches. It can be observed that our approach has an accuracy of approximately 4\% higher than the baseline. Moreover, we find that the data is polarized, and the approach with reputation-weighted node selection has a significant improvement over the one without. The impact of the data filtering module on accuracy is not significant, but this does not mean that our data filtering module is meaningless. We will analyze it in the following.
    
    Fig. \ref{fig:ablation_var} shows the variance of data obtained by different approaches. It can be found that the variance of our approach is much lower than that of the baseline. Moreover, our approach still has a lower variance than the conventional approach even with some sub-algorithms (node selection and data filtering) removed. This indicates that both of our proposed sub-algorithms can effectively improve data quality, but they work in different mechanisms. The data filtering algorithm filters more concentrated data by sliding time windows, while the reputation-weighted node selection algorithm filters out malicious nodes to reduce error data and improve data quality. They can work together to provide higher-quality data for oracles.
    
    Fig. \ref{fig:ablation_time} shows the time of data obtained by different approaches. Both our approach and the approach with some sub-algorithms removed (node selection and data filtering) can effectively reduce the response time of oracle tasks. However, it can be found that the filtering algorithm can significantly reduce response time than the reputation algorithm.
    
    The three experiments in Fig. \ref{fig:ablation} demonstrate the performance advantage of our approach over the baseline. This indicates that our designed reputation and data filtering algorithms are effective, and they work in different directions. The reputation algorithm can improve the accuracy of the data more significantly, while the data filtering algorithm can reduce the data acquisition time more effectively.

    \subsection{Security}
    
    \begin{figure}[]
    \centering
    \subfloat[Accuracy of data]{\includegraphics[width=0.5\linewidth]{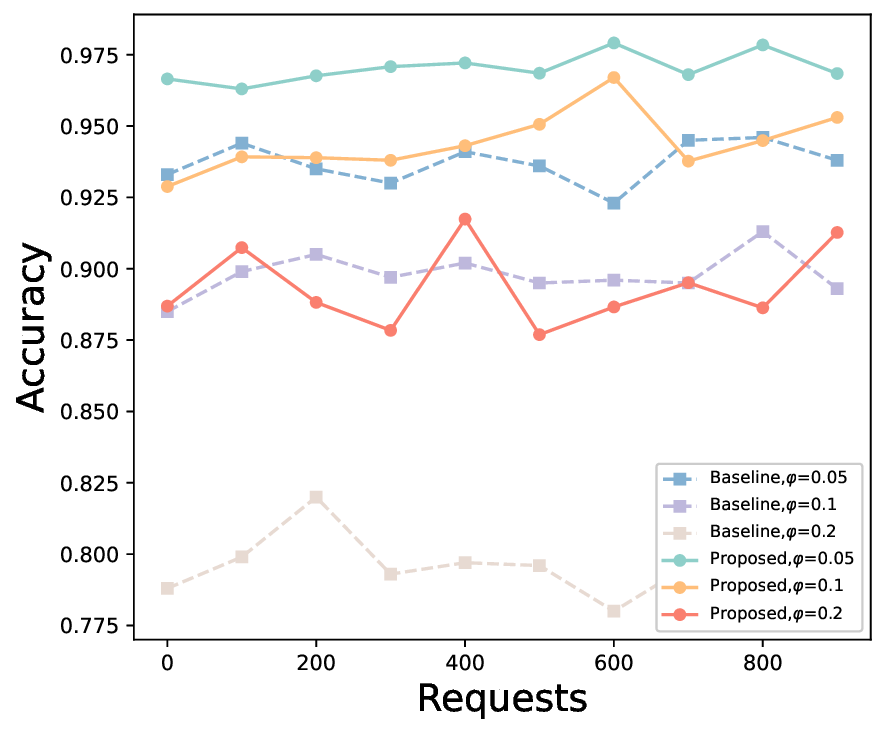}%
    \label{fig:err_rate_acc}}
    \hfil
    \subfloat[Variance of data]{\includegraphics[width=0.5\linewidth]{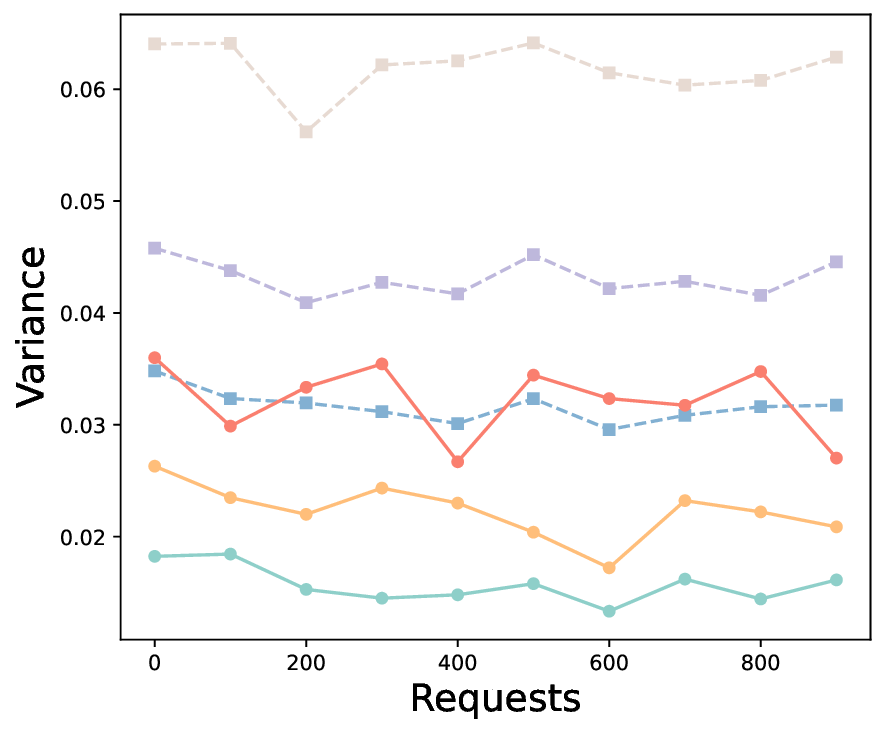}%
    \label{fig:err_rate_diff}}
    \caption{The accuracy and variance of the data obtained by each scheme when there are different proportions $\varphi$ of [5\%, 10\%, 20\%] of the malicious nodes.}
    \label{fig:err_rate}
    \end{figure}
    
    \begin{figure}[b!]
    \centering
    \subfloat[Reputation of node]{\includegraphics[width=0.5\linewidth]{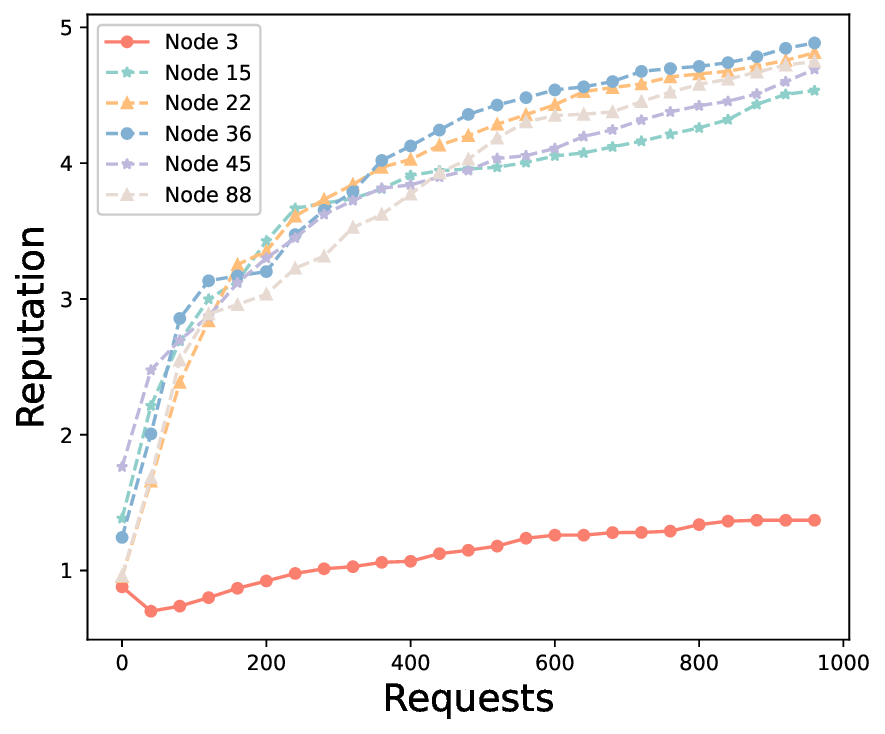}%
    \label{fig:node_reputation}}
    \hfil
    \subfloat[Selection time of node]{\includegraphics[width=0.5\linewidth]{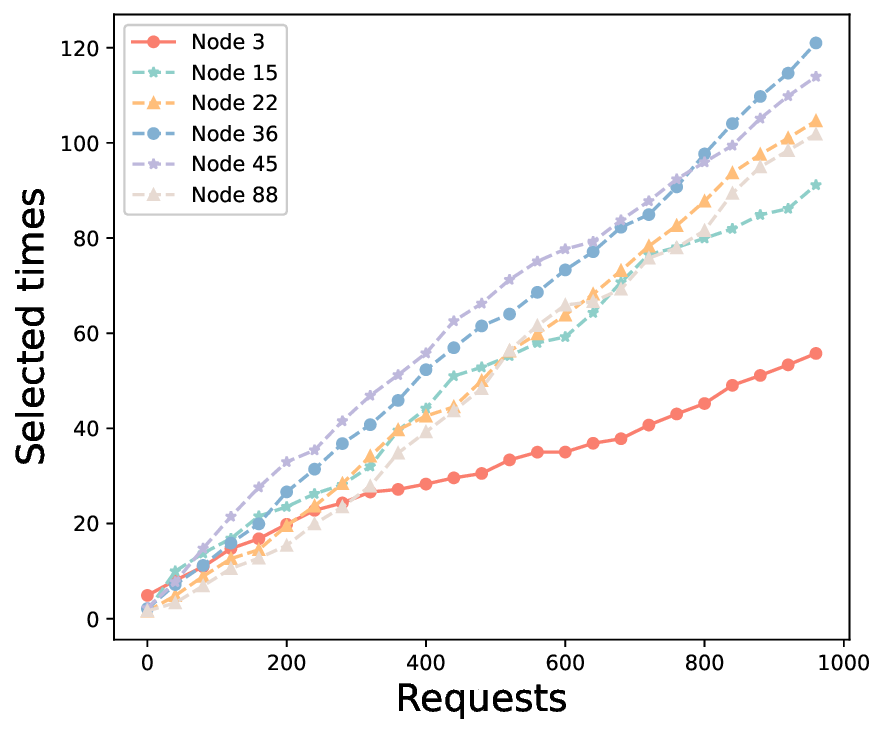}%
    \label{fig:node_select_time}}
    \caption{Change of reputation and selection times of randomly selected 6 nodes.}
    \label{fig:node_select}
    \end{figure}

    To demonstrate the security of our proposed approach, we conduct 1000 oracle tasks with $\varphi$ set to [5\%, 10\%, 20\%], respectively. As shown in Fig. \ref{fig:err_rate}, we analyze the differences between our proposed approach and the baseline in terms of data accuracy and variance.
    
    Fig. \ref{fig:err_rate_acc} depicts the accuracy of data acquisition for different approaches. It can be observed that our approach consistently outperforms the baseline in terms of accuracy, even under the same $\varphi$. As the $\varphi$ increases, the accuracy of data acquisition decreases, but our approach shows a slower decline. This indicates that our approach can provide high-accuracy data in oracle networks with varying proportions of malicious nodes, and the advantage becomes more pronounced as the number of malicious nodes increases.
    
    Fig. \ref{fig:err_rate_diff} shows the variance of data acquisition for different approaches. Similar to the accuracy, our approach consistently maintains lower data variance than the baseline. As the $\varphi$ increases, the variance of data acquisition for all approaches increases, but our approach exhibits a significantly slower growth rate. Even when 20\% of the nodes are malicious, our approach can still maintain a similar level of variance as the baseline when only 5\% of the nodes are malicious.
    
    The two experiments in Fig. \ref{fig:err_rate} demonstrate that our proposed approach can provide trustworthy high-quality off-chain real-time data for blockchain, even as the number of malicious nodes in the oracle network increases.

    \subsection{Reputation Update and Node Selection}
    
    \begin{figure*}[!t]
    \centering
    \subfloat[Accuracy of data]{\includegraphics[width=0.33\linewidth]{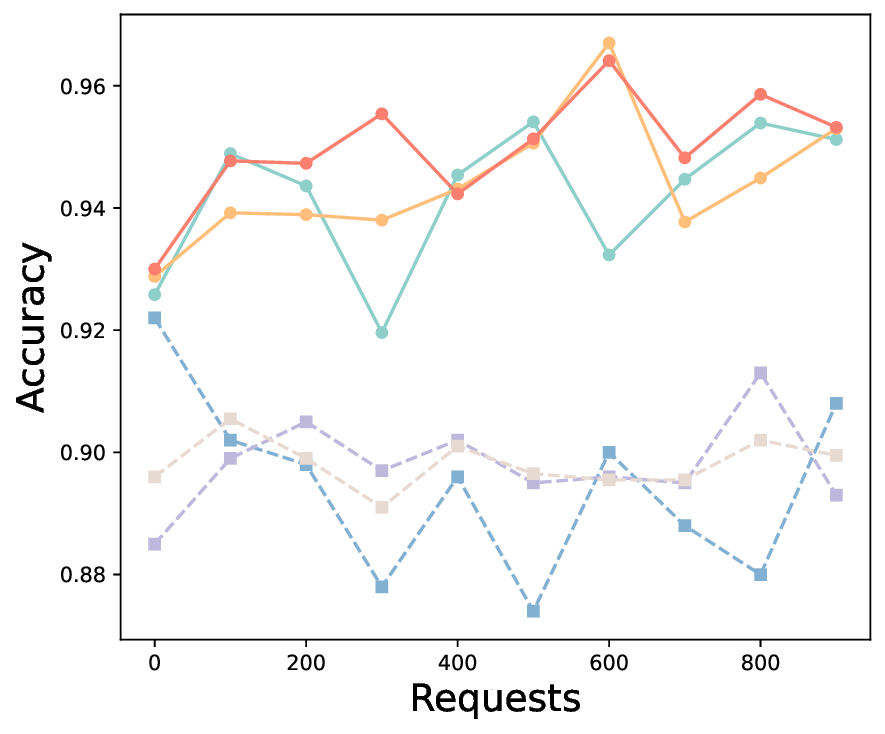}%
    \label{fig:T_acc}}
    \hfil
    \subfloat[Variance of data]{\includegraphics[width=0.33\linewidth]{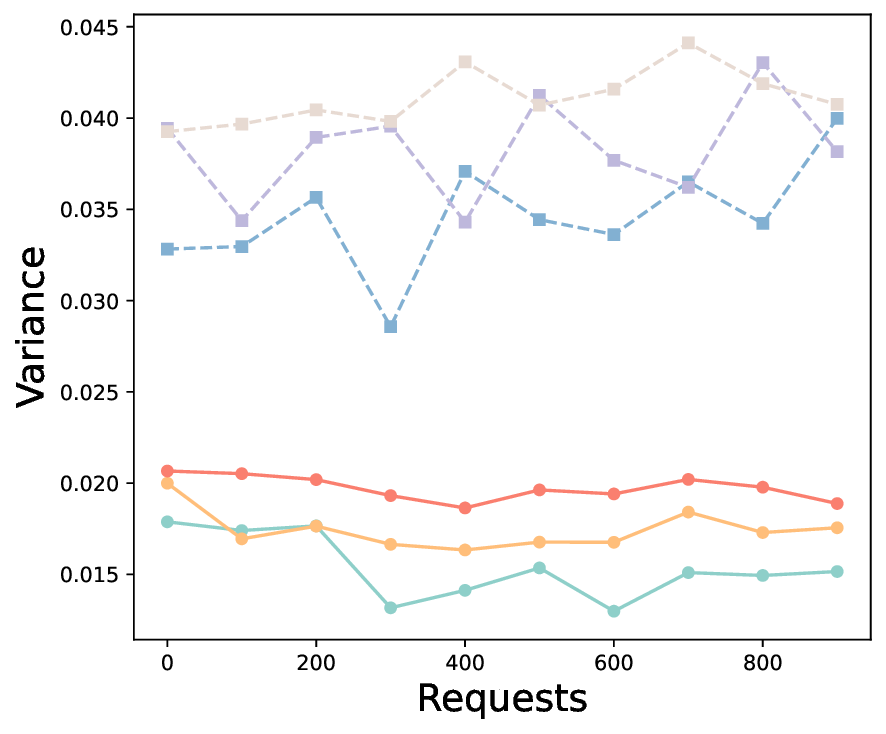}%
    \label{fig:T_var}}
    \hfil
    \subfloat[Time of obtaining data]{\includegraphics[width=0.33\linewidth]{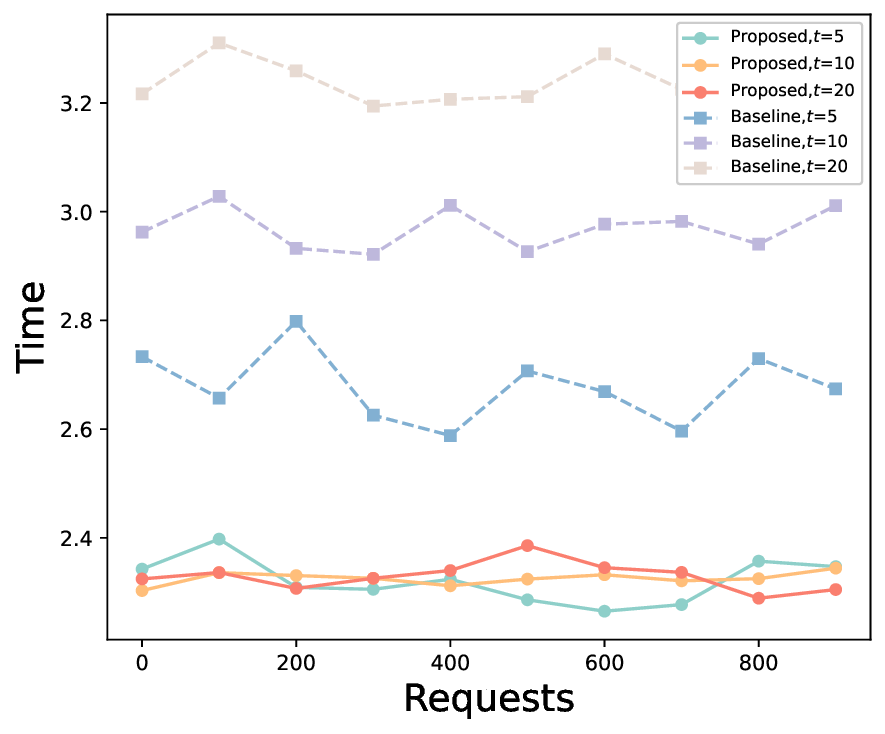}%
    \label{fig:T_time}}
    \caption{The relationship between the number of selected nodes $t$ and the accuracy, variance, and time of data acquisition.}
    \end{figure*}

    To demonstrate the effectiveness of the reputation mechanism and node selection algorithm proposed in this paper, we randomly select six oracle nodes (Node3 is a malicious node) to observe their reputation value changes and the number of nodes selected.

    Fig.\ref{fig:node_reputation} shows the changes in the reputation values of the nodes. It can be observed that, except for the malicious node (Node3), the reputation values of the other nodes improved as they honestly participated in the tasks. The reputation values increase rapidly in the initial stages and then gradually level off as the number of iterations increases, which is consistent with our goal of avoiding the Matthew effect. For malicious nodes, the rate of increase in reputation values is much slower compared to honest nodes.
    
    Fig.\ref{fig:node_select_time} presents the frequency of node selections. Similar to the reputation update pattern, the malicious node is rarely selected due to its lower reputation value. We can observe that the malicious node still has some participation opportunities since we ensure that $1 \leq R_i$ in our simulation. However, in a real production environment, we recommend imposing stricter constraints on malicious nodes to ensure the security and credibility of the oracle network.
    
    \subsection{Scalability}

    To explore the scalability of this scheme, we adjust the number of selected nodes $t$ and analyze the impact of $t$ increase on the performance of the oracle.

    Fig.\ref{fig:T_acc} shows the relationship between the node's data accuracy and $t$. We can observe that the impact of $t$ on accuracy is not significant.
    
    Fig.\ref{fig:T_var} shows the relationship between the node's data variance and $t$. In contrast to accuracy, $t$ has a significant impact on the data variance. The smaller the value of $t$, the smaller the variance of the obtained data, which indicates higher data quality. However, since $t$ is the threshold for returning data, a smaller $t$ may result in lower security. Therefore, $t$ should be set based on the actual scenario's requirements.
    
    Fig.\ref{fig:T_time} shows the relationship between the node's data acquisition time and $t$. We can see that as $t$ changes, the time taken for baseline data acquisition increases rapidly, while our proposed approach can maintain a relatively stable time. We have also analyzed the impact of our node selection and data filtering process on data acquisition time. This is because our node selection algorithm selects nodes that can obtain data more quickly, and our data filtering algorithm filters data that is more concentrated in time.

    The above experiments show that the proposed scheme has good scalability. As the number of selected nodes $t$ increases, the accuracy and response time of the obtained data can remain relatively stable.

    \subsection{Robustness}

    \begin{figure}[]
    \centering
    \subfloat[Accuracy of data]{\includegraphics[width=0.5\linewidth]{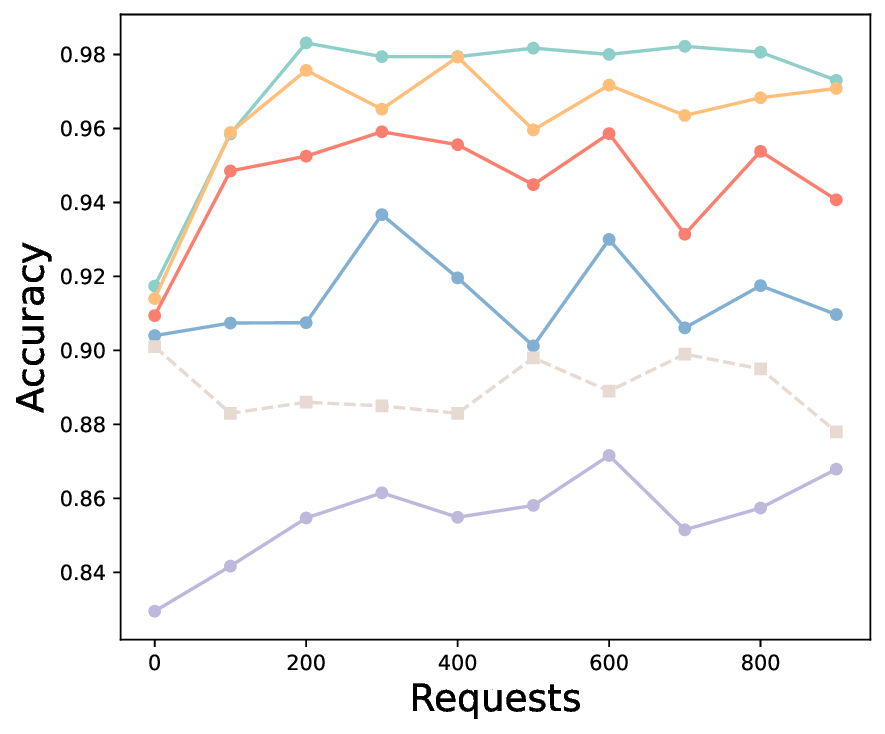}%
    \label{fig:Alpha_acc}}
    \hfil
    \subfloat[Time of obtaining data]{\includegraphics[width=0.5\linewidth]{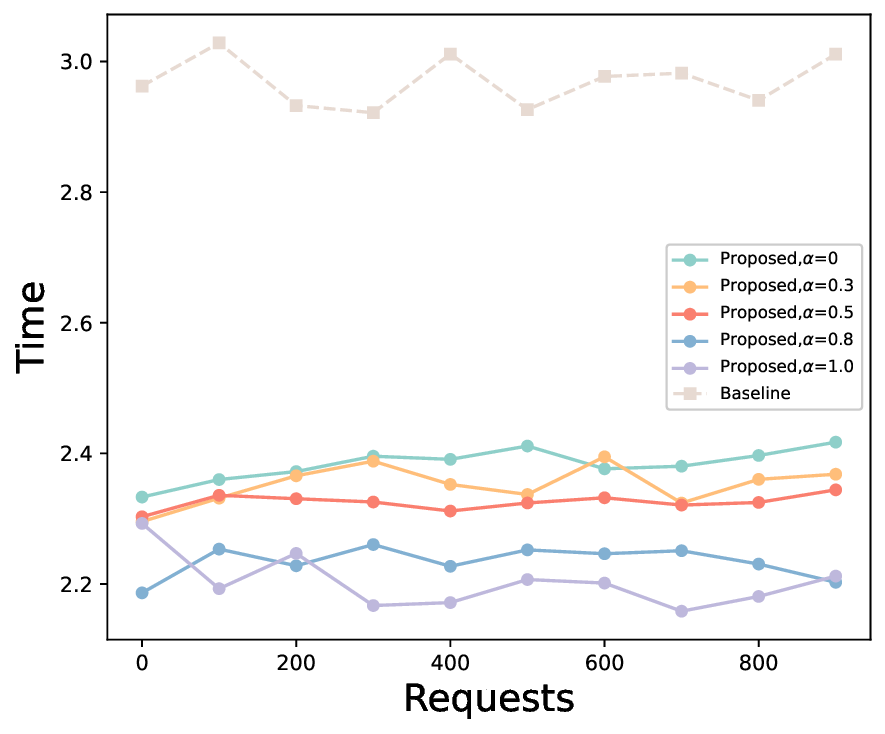}%
    \label{fig:Alpha_time}}
    \caption{The relationship between the hyperparameter $\alpha$ and the accuracy and time of the obtained data.}
    \label{fig:alpha}
    \end{figure}

    \begin{figure}[]
    \centering
    \subfloat[Accuracy of data]{\includegraphics[width=0.5\linewidth]{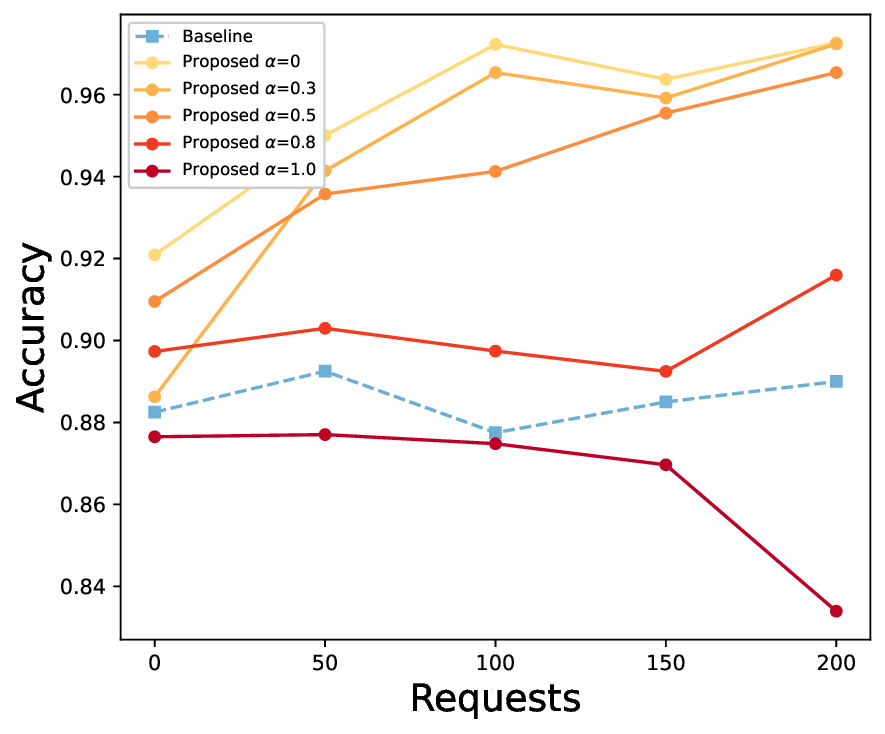}%
    \label{fig:eth_acc}}
    \hfil
    \subfloat[Time of obtaining data]{\includegraphics[width=0.5\linewidth]{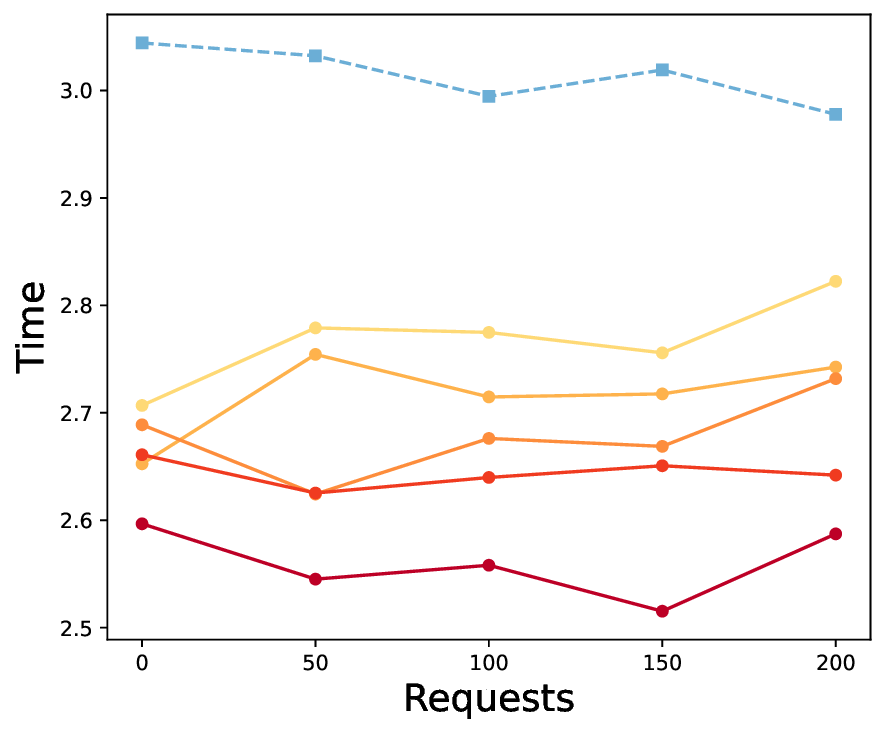}%
    \label{fig:eth_time}}
    \caption{Availability on a real blockchain.}
    \label{fig:eth}
    \end{figure}

    Fig.\ref{fig:Alpha_acc} shows the relationship between the data accuracy of the node and the hyperparameter $\alpha$. We can see that $\alpha$ has a significant impact on data accuracy and a smaller $\alpha$ leads to higher data accuracy. This is because $\alpha$ controls the balance between data accuracy and reputation computation time, and a smaller $\alpha$ increases the probability of selecting nodes with high data accuracy. However, we also find that when $\alpha$ is set to 1.0, which means that the reputation calculation only considers the time to obtain data rather than the data accuracy, the data accuracy of our scheme is worse than the baseline.

    Fig.\ref{fig:Alpha_time} presents the relationship between the node's data acquisition time and $\alpha$. As $\alpha$ increases, the time taken by our scheme to acquire data decreases gradually. Regardless of the changes in $\alpha$, our scheme can achieve a lower response time than the baseline. This is consistent with our original design intention and previous verification, as the data filtering algorithm can reduce the data acquisition time.

    Finally, to verify the availability of the proposed scheme on the real blockchain, we use Truffle Suite to generate an oracle network composed of 100 active Ethereum blockchain nodes and deploy smart contracts on the local Ethereum blockchain. We perform the same experiment as Fig. \ref{fig:alpha}, and the results are shown in Fig. \ref{fig:eth}. The results are similar to Fig. \ref{fig:alpha}, which shows that the proposed scheme is still available in real blockchain scenarios.

\section{CONCLUSION}
\label{conclusion}
    This paper proposes a secure and trustworthy oracle scheme to address the issue of off-chain data in the IIoT. We design a novel node selection algorithm based on VRF and reputation mechanism to anonymously select high-quality nodes while improving the quality of off-chain real-time data. Moreover, we propose a sliding window-based data filtering algorithm to improve data consistency and data acquisition efficiency. We demonstrate that the proposed scheme outperforms the baselines via security analysis and simulation experiments. Nonetheless, further research is required to address some of the open issues associated with our approach, such as identifying and handling malicious nodes, and mitigating the impact of data missing and noise.

\bibliographystyle{IEEEtran}
\bibliography{IEEEabrv,myref}

\begin{IEEEbiographynophoto}{Peng Liu}
received his Ph.D. degree in 2017 from Beihang University, China. He began his academic career as an assistant professor at Guangxi Normal University in 2007 and was promoted to full professor in 2022. His current research interests are focused on federated learning and blockchain.
\end{IEEEbiographynophoto}

\begin{IEEEbiographynophoto}{Youquan Xian}
received the BEdegree from the BeiBu Gulf University, in 2021. He is currently working toward the master’s degree at Guangxi Normal University. His research interests include blockchain, edge computing, and federated learning.
\end{IEEEbiographynophoto}

\begin{IEEEbiographynophoto}{Chuanjian Yao}
received the BEdegree from the Guilin University of Electronic Technology, in 2020. He is currently working toward the master’s  degree  at Guangxi Normal University. His research interests include blockchain, edge computing, and deep reinforcement learning.
\end{IEEEbiographynophoto}

\begin{IEEEbiographynophoto}{Peng Wang}
received his master's degree from Guilin University of Technology in 2018. He is currently working toward a doctor's degree at Guangxi Normal University. His research interests include blockchain, data fusion, and data security.
\end{IEEEbiographynophoto}

\begin{IEEEbiographynophoto}{Li-e Wang}
 received the Ph.D. degree from Guangxi Normal University, Guilin, in 2022 and the M.S. degree from Hunan University, Changsha, in 2007. Now, she is a professor in School of Computer Science and Engineering, Guangxi Normal University, Guilin, China. Her research interests mainly include data privacy, recommendation, distributed system security and machine learning. She has authored more than 40 refereed papers in these areas. She has served as a reviewer for several high impact research journals and ACM/IEEE flagship conferences.
\end{IEEEbiographynophoto}

\begin{IEEEbiographynophoto}{Xianxian Li}
received the Ph.D.degree from the School of Computer Science and  Engineering,  Beihang University, Beijing, China, in 2002. He worked as a professor at Beihang University during 2003-2010. He is currently a professor with  the School of Computer Science and Information Technology, Guangxi Normal University, Guilin, China. His research interest includes information security.
\end{IEEEbiographynophoto}

\end{document}